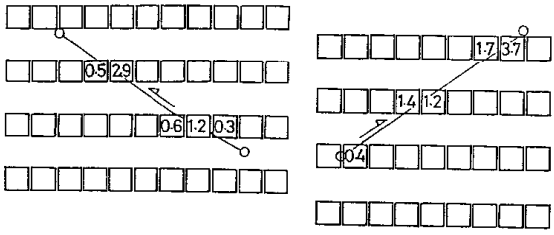

Ev. No. 2099

Fig. 1a

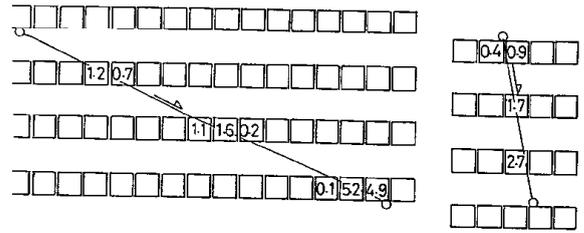

Ev. No. 165-79

Fig. 2a

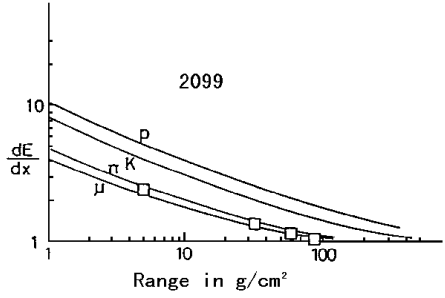

Fig. 1b

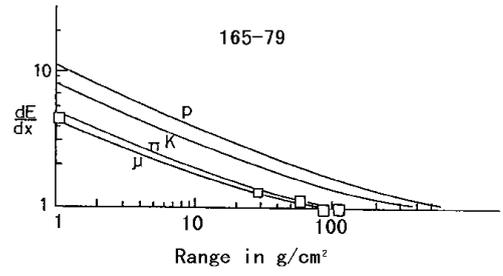

Fig. 2b

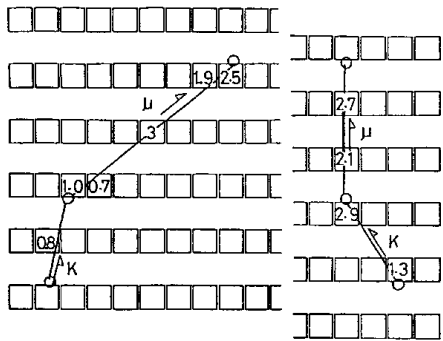

Fig. 3a    867

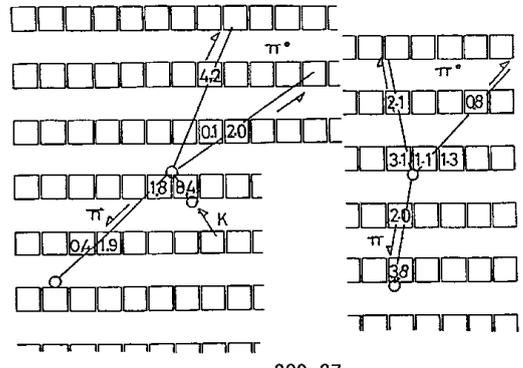

229-37

Fig. 4a

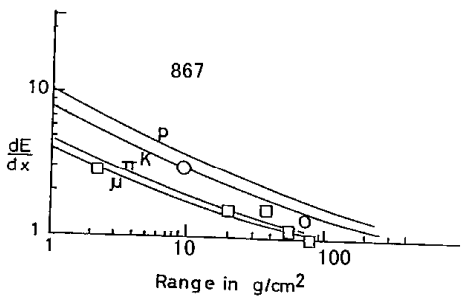

Fig. 3b

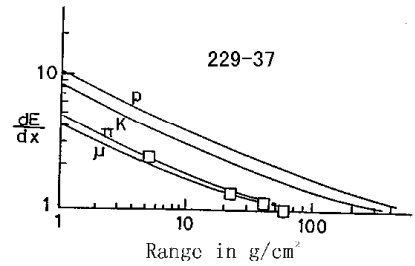

Fig. 4b

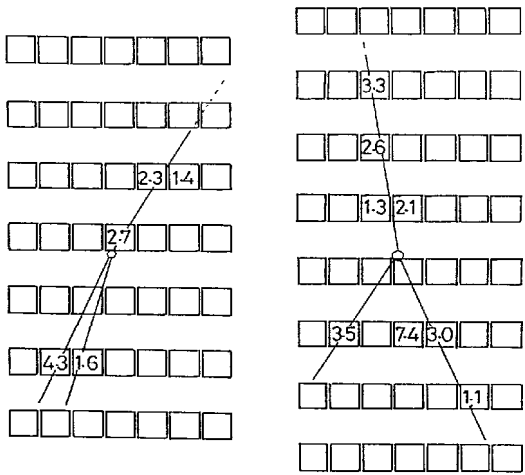

Fig. 5

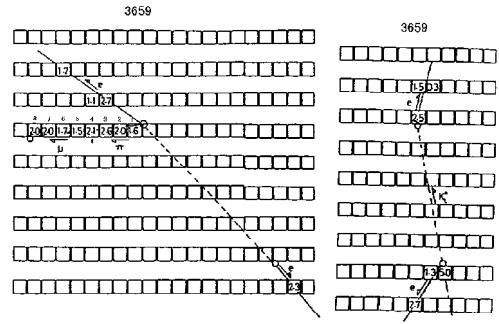

Fig. 6 a

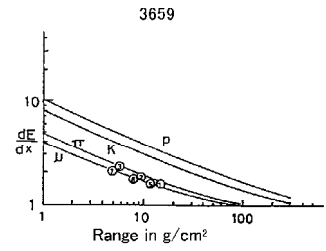

Fig. 6 b

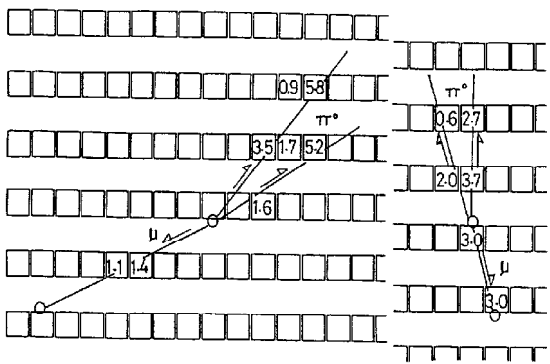

Fig. 7

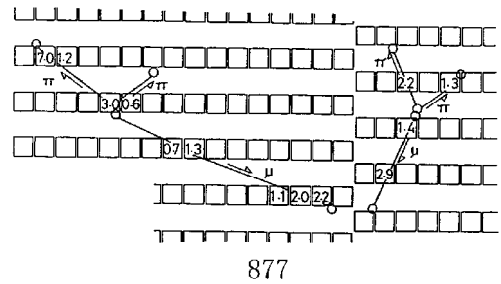

Fig. 8a

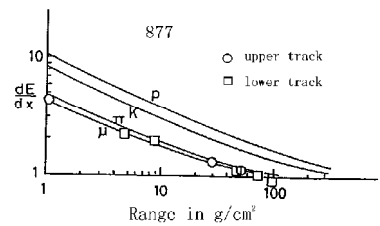

Fig. 8b

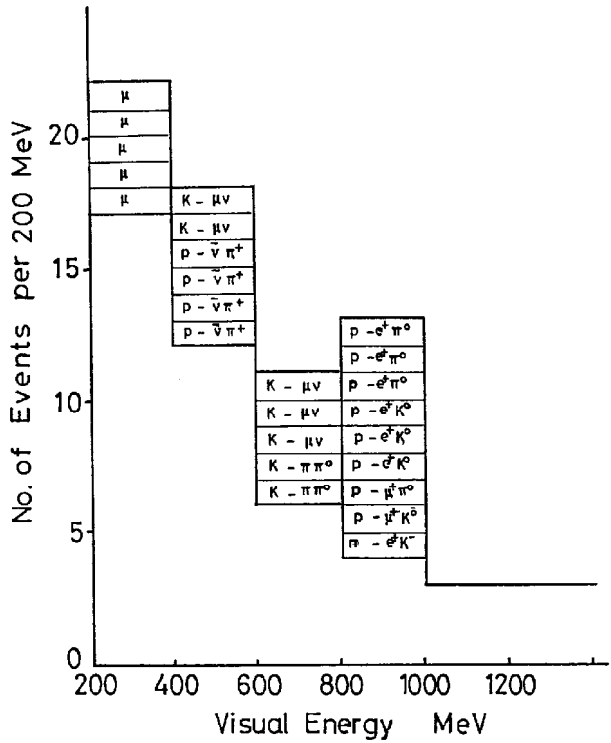

Fig. 9.

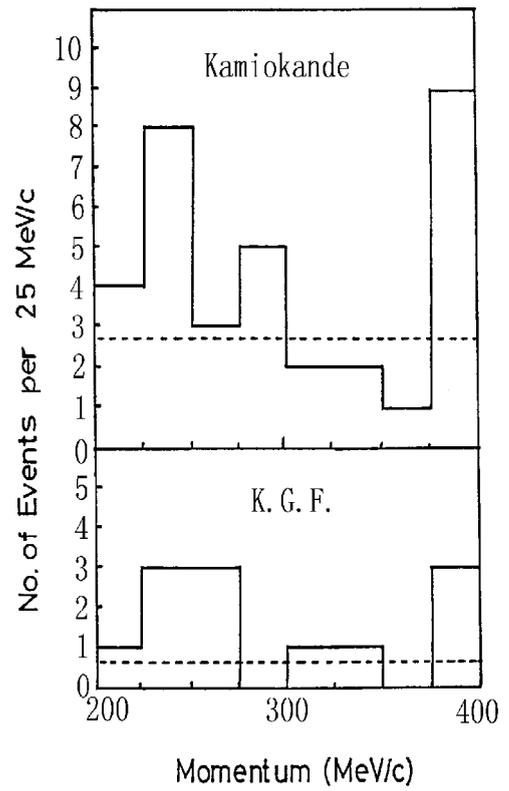

Fig. 10

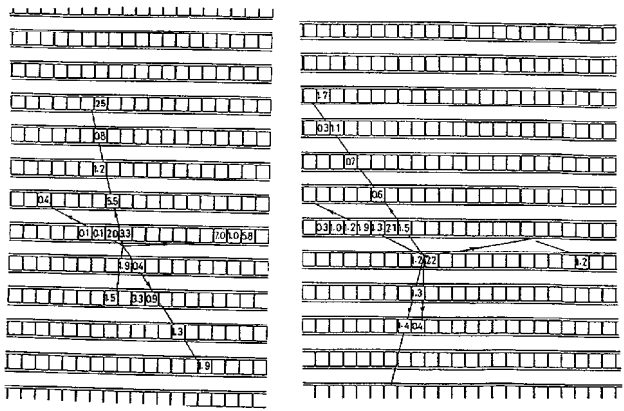

Fig. 11

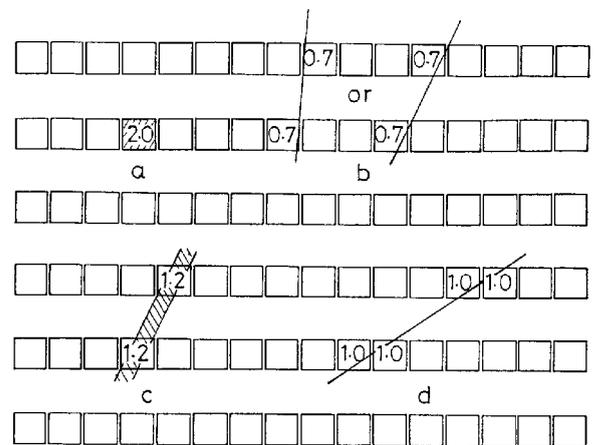

Fig. 12

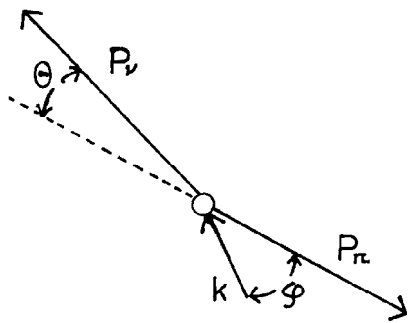

Fig. 13

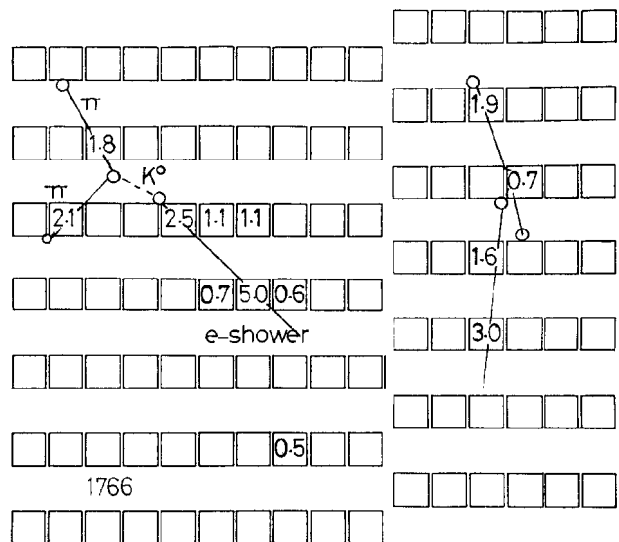

Fig. 14a

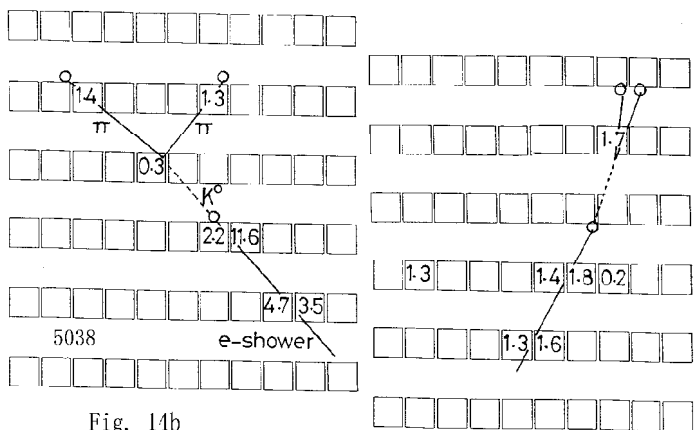

Fig. 14b

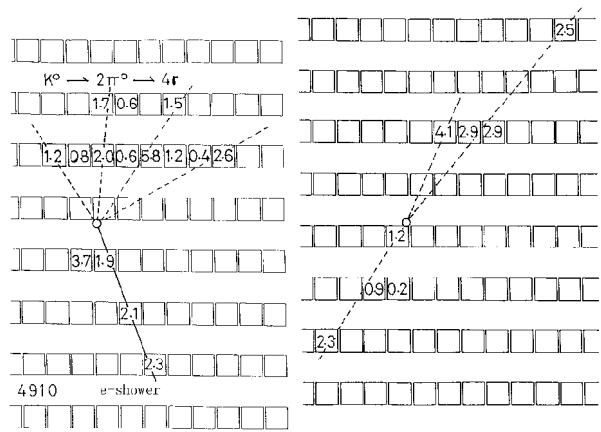

Fig. 14c

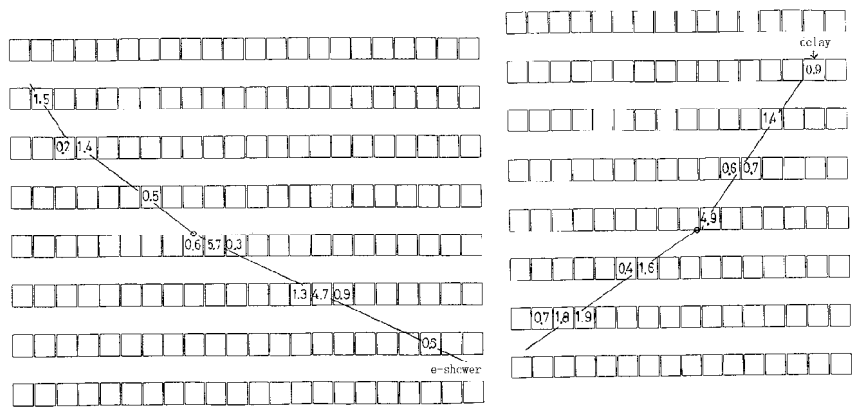

64-125

Fig. 14d

# EXPERIMENTAL EVIDENCE FOR G.U.T. PROTON DECAY

H.Adarkar, S.R.Dugad, M.R.Krishnaswamy, M.G.K.Menon, and B.V.Sreekantan

Tata Institute of Fundamental Research, Bombay, India

Y.Hayashi, N.Ito, S.Kawakami, S.Miyake and Y.Uchihori

Osaka City University, Osaka, Japan

Abstract

Deep underground in Kolar Gold Fields, in southern India, an experiment to detect proton decay had been carried out since the end of 1980. Analysis of data yielded the following results; (l) the life time of proton is about $1 \times 10^{31}$ years, (2) it decays into wide spectrum of decay modes, $p \rightarrow e^+ \pi^0$, $p \rightarrow \bar{\nu} K^+$ and so on, and (3) the life time and the distribution of decay modes are close to the predictions of SU(5) SUSY GUT. Four events representing possibly neutron oscillation are also seen.

I. Introduction

Kolar Gold Fields(KGF) in southern India have ideal conditions for underground experiments, due to relatively flat topography and uniform distribution in the characteristics of rock within a wide range of area and depth. A series of experiments underground has been carried out deep underground in KGF since 1960. The first experiment was performed to observe the Depth-Intensity relation for high energy cosmic ray muons and observations were made at several depths from 816 to 8400 hg/cm$^2$ (Ref.l). Observations at the deepest depth of 8400 hg/cm$^2$, which showed no count during a period of 2 months in a 3 m$^2$ wide angle detector, gave the first suggestion about the lifetime of proton being longer than $3 \times 10^{28}$ years (Ref.2), and indicated the possibility of detecting natural high energy neutrinos. In 1965, the first cosmic ray neutrino was observed in KGF by the Tata-Osaka-Durham group (Ref.3). The results obtained from this study agree well with the calculated intensity of cosmic ray neutrinos produced in the atmosphere as decay products of kaons, pions and muons. Though general features of cosmic rays observed deep underground were known and understood from these experiments, there were still some rare events observed which were difficult to understand in terms of known phenomenon. These rare events ( Ref.4 ) are (a) 'Kolar events' which may be due to some massive long lived particle, and (b) 'Anomalous Cascade Showers' showing double cores with evidence for large transverse momentum. However, their energies are very high, for example, several GeV for (a) and a few hundred GeV for (b), and their rate of occurrence has been observed to be very small. Therefore, these rare events do not contribute any significant background for possible detection of proton decays. Since the flux of magnetic monopole has also been shown to be undetectable level at KGF (Ref.5) , only neutrino interactions offer a background for the proton decay phenomena.

The problem of nucleon instability has attracted considerable interest since l974 motivated by the development of the Grand Unification Theory (GUT ) (Ref.6 ) in which strong, electromagnetic and weak interactions were described by means of a single Gauge theory. Certain models, (Ref. 7), of GUT have suggested possible violation of the law of baryon number conservation. These theories predict a number of processes with non-conservation of the baryon number B which result in a finite lifetime for the nucleon. Since the mass of Gauge boson (X-particle) in GUT is to be of order $10^{15}$ - $10^{16}$ GeV, an experimental proof of such theory is accessible only through an experimental detection of non-conservation of baryon number. One of the main processes is the decay of protons, for instance, through one of the probable channels p → $e^+ \pi^0$ with $\triangle B$ = -l. In some models, the process of neutron-antineutron oscillation, related to a change in the baryon number $\triangle B$ = -2, is also predicted. The estimated range for the nucleon lifetime is given to be $10^{30-33}$ years. A similar value has been estimated for the n-n~ transition time also for neutrons within nuclear matter (Ref. 8). Therefore, an experimental study on nucleon instability is now of fundamental and immediate interest, aiming to verify the validity of GUT through detection of the products from nucleon decays.

We have made a careful check of earlier KGF data recorded deep underground and found a few events which may be explained as representing nucleon decay phenomena though the detector was not proper to study and analyze such events in detail. From these observations we estimated the lifetime of the nucleon to be of the order of 3 x $10^{30}$ years. Therefore, we developed a new design for a detector for the proton decay experiment in 1979. This 'Phase I' detector which comprised of 1600 proportional counters having a total weight of 140 tons of iron, was installed deep underground at the depth of 7000 hg/cm$^2$ and started making observations on possible proton decay phenomenon at the end of 1980.

After few years of operation, preliminary results showing candidate events for proton decay were published (Ref. 9; further publications are also listed here). As the observed rate for possible candidate events for proton decay in the phase I detector was smaller than our expectation from earlier observations, a new and larger 'Phase II' detector was installed at a depth of 6045 hg/cm$^2$ (Ref. 10). This detector became operational in 1985. The Phase II detector comprised of about 4000 proportional counters and total weight of 260 tons.

Since the management of KGF mines decided to start closing the mines in 1993, the proton decay experiment stopped data taking in 1993. However, we have accumulated a large database with our experiments, Phase I and II by combining all data obtained over a total period of 13 years, to enable us discuss the question of proton lifetime in some detail. Nevertheless, it is not enough for estimating the branching ratios in detail for various decay modes which are necessary to distinguish between various theoretical ideas. Our results, as discussed later, show that there is a wide spectrum of decay modes and that all the observed decay modes seem to have more or less similar weight. Although the statistics is poor, there is a general trend that kaons are observed more frequently among the products of proton decay.

A number of other experiments (Ref.11) have also looked for starting proton decay after 1982, (for example, Mont Blanc experiment by an Italian group, the IMB experiment by an American group, the Kamiokande by the Japanese group, and the Frejus experiment by a French-German collaboration). The present consensus among these other experiments seems to be that they have not found any evidence for proton decay yet, and that the lower limit on the lifetime of proton is of the order of $10^{33}$ years (Ref. 12). Contrary to this conclusion, as shown in our earlier reports (Ref. 9), our result is that protons are decaying with lifetime of about $10^{31}$ years. The apparent contradiction between these conclusions does not mean a complete disagreement between the observations. Actually, some of the observations in other experiments are quite similar to our observations. A detailed discussion on the observational situation is presented later in the section on Discussion. Our positive result for the existence of the proton decay phenomena may not be crucial until other experimental groups also reach similar conclusions. A clue to this question comes from the interpretation the experimental results.

II. Experimental Details

The Phase I detector, comprised about 1600 proportional counters, with a floor area of 4 x 6 $m^2$. Proportional counters of 10 x 10 $cm^2$ cross sectional area, 2.3 mm thick iron walls and 4 m or 6 m length were placed in alternate horizontal layers in a crossed layer configuration. Iron plates of l.2 cm thickness were inserted in between the layers. There were total 34 layers with a total weight of the detector about 140 tons.

The Phase II detector was also of similar type but used only 6 m long proportional counters. The floor area was 6 x 6 $m^2$ and there were a total of 60 counter layers in the detector. One major difference was in the thickness of the iron plates placed between the counter layers which only 0.6 cm in the Phase II detector. The total weight of this detector was about 260 tons.

Phase I and II detectors were installed at the depths of 7000 $hg/cm^2$ and 6045 $hg/cm^2$ respectively. The observed rate of atmospheric muons passing through these detectors was approximately 2 and 11 per day respectively.

A trigger was generated when one of the following conditions satisfied;
(1) a 5-fold coincidence between the layer pulses from consecutive 11 layers, or
(2) 2-fold coincidence between consecutive 3 layers for the layer pulses generated by signals from more than 2 counters in the layer.

Data including the address. pulse height and timing of each fired counter was recorded for every event. In order to calibrate the counters, pulse height distribution for each counter is also recorded periodically. In the pulse height distribution which is mainly due to gamma rays from the radioactivity of surrounding rock, there is a sharp peak at around 6.4 keV due to the photoelectric effect in iron atom induced by gamma rays. This is a characteristic X-ray line of iron. Using this peak for the calibration, the energy loss of 20 keV corresponding to the traversal of 10 cm path length by relativistic particle in the proportional counter is used as a unit to measure pulse heights.

Although the triggering rate is few per hour, the records of accidental coincidence due to gamma rays are easily distinguished by looking at the pattern of fired counters. Picking up events of interest is very simple because the event rate is so low. For example, in Phase I experiment, the rate of atmospheric muons penetrating the detector was only about 2/day, the rate of neutrino induced muons from rock with large zenith angles was only about l/month, the rate of showers entering the detector from outside was several per year and the rate of confined events which include neutrino interactions and proton decay candidates was also only several per year.

Analysis of events of interest is based on a pattern of fired counters and ionization measured in each counter. The direction of motion of a stopping track is determined as the direction of increasing ionization (dE/dx). Figure 1b shows plot of the ionization versus the range of the track. As seen from this figure, such a plot can be used for identification of the particle. Although muons and pions are difficult to distinguish, kaons and protons are easily separated from the muons. The last point, which gives the highest ionization along the track, is also used to determine the range of track in the final iron plate.

The estimate of the energy of a track is obtained in two ways; (l) from the range-energy relation for the track and (2) from the calorimetric method; total ionization (in units of 20 keV energy loss) multiplied by a characteristic energy, which is 23 MeV and 15 MeV for Phase I and Phase II detectors respectively, plus the mass energy of the particle. Therefore, measurement of ionization is very important for our analysis. It is one of the main reasons that counters of 10 x 10 $cm^2$ cross-sectional area were preferred for use in our experiments. The advantages are; (1) the error in the ionization measurement in each counter is less than 20 % and (2) it reduces the gap effect which, in the present case, is less than 7 % on the average, including the thickness of the iron wall. In some cases, however, the gap effect is quite large for vertical tracks at special positions, but it is also useful in such cases as it gives a very accurate measurement of the position of the track.

In case of electro-magnetic cascade showers, the direction of motion is rather easy to find from the pattern of fired counters, and the energy is estimated by the calorimetric method. The energy calibration and the error in energy determination have been obtained from shower simulation using the Monte Carlo program "GEANT". It has been seen that the energy estimate using the calorimetric method, mentioned above, agrees well with the results from simulations and the error in energy determination is estimated to be about 20 % at 1 GeV.

The observational requirements for the nucleon decay phenomenon are, total energy E = 938 MeV, momentum sum P = 0, and the change of the baryon number B and the lepton number L should be equal ($\triangle$B = $\triangle$L). However, these are modified by nuclear effects and secondary interactions and also due to errors in measurements. Since there is no unique way to define a candidate event, the considerations are different on a case by case. It is to be noted that the requirements on total energy and momentum include contributions due to neutral particles which may not be observed; therefore, presence of a neutrino is usually assumed in cases of one-prong events or other well-known decay phenomena.

Though our detector has the same sensitivity right up to the edge of the detector, a certain detector volume is required around the vertex of an event for it to be fully confined within the detector. For defining this confinement fiducial volume, detector volume within 30 cm from all outside edges has been excluded for selection of the vertices of candidate events. With these criteria for confined events, the fiducial volumes for Phase I and Phase II detectors are 80 and 180 tons respectively. Of course, there may be some events which are fully confined even when their vertices are outside the fiducial volume and there may be few events which may not be fully confined even with their vertices inside the selected confinement volume. Therefore, independent of this fiducial volume, we analyze all observed events and some events are rejected due to the edge effect only after detailed considerations.

III  Results

The total exposure factors for the two detectors, calculated by multiplying the fiducial volume by operational time in units of kilo-ton years ( kty ) are as follows,

Phase  I    0.675 kty    Phase II    0.995 kty    Total exposure    l.67 kty

Out of about 100 confined events observed during this exposure, 84 events have been analyzed as having more than 200 MeV in visible energy. Among these events, the proton decay candidates are as follows;

(l)  p $\rightarrow \bar{\nu} \pi^+$

Figures la and 2a shows the candidate events for this decay mode observed in Phase I and II detectors respectively. Figures 1b and 2b show the relation of ionization (dE/dx) vs. residual range for these events.  It is seen that experimental points are fitting well with the pion (and muon) curves, and the direction of motion is clearly towards larger ionization. Four such clear events belonging to this category have been observed as listed below;

| Event No. | (Ix Ec)+mc$^2$ | Range(g/cm$^2$) | Energy (MeV) |
|---|---|---|---|
| 2099   (I) | 460 MeV | 172 ± 16 | 450 ± 24 |
| 165-79  (II) | 463 | 171 ± 10 | 447 ± 15 |
| 775-70  (II) | 433 | 185 ± 20 | 469 ± 30 |
| 1807-14  (II) | 484 | 163 ± 15 | 435 ± 22 |

The table above shows the energy estimates from both methods, range and calorimetric. Of course, the range method is more reliable and accurate than ionization.

It is expected that the energy of the pion in this decay mode is about 480 MeV in the case of a free proton. But it may be reduced to about 460 MeV by the nuclear effects, with a half-width of about 50 MeV(± 25 MeV) due to Fermi energy within iron nucleus. (see Appendix (?)) The four candidate events listed above are all in this region. However, there are two types of inefficiencies; (l) about half of the cases may be missed due to nuclear interaction within the parental iron nucleus or with another nucleus in their path within iron plates, and (2) geometrical inefficiency due to the length of the event which is about 1 to 1.5 m in these detectors. Since the average densities of Phase I and II detectors are 1.7 and 1.1 g/cm$^3$, respectively, there is a large chance for the pion (muon) to go out of geometry of the detector. The background for these types of events is due to muons from neutrino CC quasi-elastic interactions and pions from NC interactions, without accompaniment of any low energy particle like a recoil proton. Considering both these processes together, the background has been estimated to be about 2 events per 100 MeV during the running time of our experiment. This result will be discussed later together with results of reported by other groups. (See Fig. 10 and Appendix)

(2)  p $\rightarrow \bar{\nu}$ K$^+$, K$^+ \rightarrow \mu^+ \nu$  or  $\pi^+ \pi^0$

In this decay mode, about half of the events look like single prong events due to K$^+ \rightarrow \mu^+ \nu$ (63 %). The range of K$^+$ track in iron is as short as about 20 g/cm$^2$ on the average. Only when the Fermi momentum for the decaying proton causes the energy of K$^+$ to be larger, the track will be able to penetrate a few layers of counters as shown in Figures 3a and 3b. Thus, independent of the existence of $K^+$ track, all single tracks, corresponding to muons of about 240 MeV/c (total energy of 260 MeV), may be considered to be the candidates of proton decay through this decay mode. More than 80 % of kaons decay after stopping and the decays in flight constitute a small fraction.   The observed candidate events for this decay mode are listed below;

| Event No. | $K^+$ energy | E(ionization) | range | E(range) |
|---|---|---|---|---|
| 867 | ~650 MeV | 287 MeV | 91 ± 10 g/cm$^2$ | 283 ± 20 MeV |
| 1574 | - | 251 | 85 ± 10 | 274 ± 15 |
| 1076-31* | ~600 | 263 | 75 ± 10 | 259 ± 15 |
| 1103-10 | ~560 | 235 | 80 ± 15 | 267 ± 20 |
| 1326-81 | ~700 | 266 | 74 ± 10 | 257 ± 15 |
| 1334-42* | ~650 | 248 | 80 ± 7 | 267 ± 10 |
| 1393-37 | - | 259 | 90 ± 14 | 282 ± 20 |
| 1400-63* | - | 247 | 77 ± 6 | 262 ± 9 |
| 1459-5 | - | 260 | 70 ± 6 | 252 ± 9 |
| 1677-161 | - | 241 | 78 ± 10 | 264 ± 15 |

The mark * means the presence of the signal due to decay electron as the detection efficiency for decay electron from muon is about 25 % throughout this experiment. The number of candidate events shown above is clearly beyond the number of background events which are expected mainly from neutrino-induced muons. The background rate is about 1 event per 25 MeV/c bin in this experiment. (See Fig.10 and App.)   Moreover, many of the candidates have reasonable signal from the stopping $K^+$ track at the origin of the track. Because of the short track length, triggering efficiency for this decay mode is somewhat lower compared with other decay modes.

In the same category of proton decay modes, there is another decay mode $K^+ \rightarrow \pi^+ \pi^0$, with a branching ratio of 21 %.   As shown in Figures 4a and 4b, experimental analysis of this type of event reasonably matches the above decay scheme.

| Event No. | Energy of $\pi^0$ | Range | Energy of $\pi^+$ |
|---|---|---|---|
| 5426* | 246 ± 80 MeV | 39 g/cm2 | 249 ± 10 MeV |
| 229-37 | 220 ± 80 | 45 | 258 ± 10 |

Both the events have signals due to a slow kaon near their vertex. The background for this event is expected to be very small because the conditions to be satisfied are so severe. The detection efficiency is also small for these low energy e.m. cascade showers. In neutrino interactions in the energy range 0.5 – 5 GeV, the dominant process is quasi-elastic scattering with single or multi-pion production while strange particle production is known to be very small. Therefore, the fact that the existence of low energy kaons is very much beyond expectation in the candidate events, suggests clearly that these events are more likely to be produced by phenomenon other than neutrino interactions. The probability for these events being due to proton decay is high.

(3)   $p \rightarrow e^+ \pi^0$

As shown in Fig. 5, there are candidate events which nicely match with the decay scheme, p

→ $e^+ \pi^0$. As can be seen in these figures, cascade showers show their special features i.e. separate fired counters, (sometimes, skipping layers of counters), broad width of showers, somewhat random pulse height in different counters, etc.. In the middle part of the event, there is a gap of a few layers which shows a delay in the start of the cascade by gamma rays by a few radiation lengths. One side of the event has a broader width to fit $\pi^0$ decay into two gamma rays with adequate opening angle for the energy of $\pi^0$. The candidate events for this decay mode are listed below;

| Event No. | E. of $\pi^0$ | Op. angle | E. of $e^+$ | Def. angle (B.to B.) |
|---|---|---|---|---|
| 4268 | 480 ± 120 MeV | ~45° | 361 ± 130 MeV | ~ 0 |
| 4910 | 611 ± 150 | ~35° | 337 ± 100 | 29° ±5° |
| 836-47 | 405 ± 130 | ~30° | 540 ± 160 | ~ 0 |

Here the deflection angle from back to back is due to Fermi motion of the decaying proton. One possibility for Event 4268 in the above list is a decay of $\omega$, but probability may be too small (8 %). The main background for these events is due to single electron production by neutrinos. However, a study of showers produced by 1 GeV electrons through Monte Carlo simulation has shown that the probability to mimic the observed features is only about $10^{-3}$ or less. The total number of e.m. cascade showers observed in this experiment for energies (1000 ± 200) MeV is only 6 including the above events. Therefore, the background for this event seems to be negligible.

(4) $p \rightarrow e^+ K^0$

There are three candidates for this decay mode ($e^+ K^0_S$). Although the $e^+$ which is observed as an e.m. cascade shower of energy of about 350 MeV is clear, the $K^0_S$ decay into 2 pions is not so clear because of their very short track length and also due to the disturbance by the nuclear interaction. The nuclear interaction of low energy pions of energy of about 200-300 MeV has a very large cross section of 400-600 mb. Actually, one of the candidates for this category (not counted in the above three), Event No. 64-125, is also a candidate for neutron decay into $e^+ K^-$. There is another type of candidate event in this category shown in Fig.6 which seems to be proton decay into $e^+$ $K^0_L$. One side of the event shows a cascade shower probably caused by an $e^+$ with energy of about 350 MeV. On the other side, $K^0_L$ travels a distance of about l.2 m with 30° deflection angle from the direction opposite to the direction of the cascade shower. Then it decays into a pion, an electron and neutrino probably. In this particular case, the pion is identified because the decay in flight of the pion is recognized from the plots of ionization versus range as shown in Fig.6b provided both the particles are traveling almost perpendicular to the counter wall within about ± 30° as estimated from the ionization in each counter. Assuming that one neutrino was emitted on the opposite side of the pion to balance the transverse momentum, the event fits well to the decay of $K^0_L$ of about 600 MeV.

One may expect for this decay mode that the phenomenon may be composed of one electron shower and a nuclear interaction with some gap in between. It is because of the large cross section of $K^0_L$ in iron about l.2 barn i.e. corresponding to about 60 - 90 cm in the detector. From this point of view, the probability to find this type of decay is estimated to be about 30 % including the decay probability at the observed distance. The background for $e^+ K^0_L$ is supposed to be very small because of the very special pattern of the event.

(5) p → $\mu^+ \pi^0$

The event shown in Fig.7 may have to be excluded due to the lack of indication for the direction of motion for the observed track. It happens sometimes due to the lack of measurement of ionization in the region of rising ionization, in case of inclined tracks when the residual range in the last iron plate is longer than 30 g/cm$^2$. Hence, there are two ways for interpreting this event, namely, (1) one pion is produced by the NC interaction of an energetic neutrino which is converted into $\pi^0$ through charge exchange process, after traveling through a few iron plates, or (2) proton decay into $\mu^+ \pi^0$. Since the observed event fits nicely with the second possibility, as a combination of a muon and a neutral pion with energies of 400 and 600 MeV respectively, it is considered as a proton decay candidate event. For the case (1) mentioned above it seems very unlikely that among the small number of observed events, there is an event with two interaction points without accompaniment of any other track and which just fits with all the requirement for a proton decay candidate. However, because of the same reason, this candidate has somewhat higher background probability of the order of 0.1. It may also be mentioned that this event has another possibility to be interpreted as neutron decay n → e$^+$ $\pi^-$.

(6) p → $\mu^+$ K$^0$

Fig.8 shows the candidate for this decay mode. In the figure, two tracks are seen, one going up and the other going down. This is rather clear and easy to interpret but there seems to be evidence for a third track. This is deduced from the observation of excess ionization in the counter overlapping with the first counter of the upward going track and the other two counters one in each view. There is considerable uncertainty in the determination of the direction and estimate of the energy for the third track. However, the characteristics of the event are clear and reasonable, and there seems to be no other way to understand this event. The inclined track going downwards is a muon whose energy of 350 ± 15 MeV as determined from its range and ionization agree well with each other. The direction of motion and the identification of the particle are clear as seen from the plot of ionization versus range. The tracks going upwards are due to a kaon decaying into two pions, with well-known kinematics. As a result, one can deduce that the Fermi momentum of the original proton as 220 MeV/c towards a direction which is almost perpendicular to the decay particles. The background expected for this event is very small.

(7) other decay modes

There are still a few events which may be explained by some of the other expected decay modes. However, they are not as clear as the candidates discussed above. Since the primary purpose of this paper is to present clear experimental evidence for the phenomenon of proton decay, these possible candidate events are not discussed in this paper.

There may be some bias against the observation of neutron decays in this experiment. However, the number of neutron decay candidates seems to be very much less compared with proton decay. In the present database, there are 4 candidates; e$^+$ $\pi^-$, e$^+$ K$^-$, $\bar{\nu}$ $\pi^0$ and $\bar{\nu}$ K$^0$, but none of them seems to present itself as a definite case. As mentioned already, the first two have possible interpretations as proton decay candidates. The interpretation for the last two events is less definite due to lack of adequate accuracy in measurements.

IV. Analysis

Since the total number of events to be analyzed and discussed is rather small, we have not fixed any special criterion to select the candidates for nucleon decay at the very beginning and we have tried to understand each event as it is. Next, we have tried to see whether the observed characteristics of an event agree with any of the expected nucleon decay mode or not. The candidate events should satisfy the requirement within the error of measurements and nuclear effect. Hence, the considerations for analysis are different on a case by case. The separation of a proton decay candidate from an ordinary neutrino event, in general, is clear because of the large differences in the characteristics between them which are generally beyond statistical fluctuation. It is rather difficult to understand the decay candidates as the tail of the distribution for neutrino events. Since the detector is not isotropic, it is difficult to measure all the required components and the quality of measurement depends on the spatial configuration of the event in the detector. Hence, there is a grading in the interpretation of events from very clear cases to somewhat ambiguous cases. Therefore only simple type of events is discussed here in this paper and the very complicated events are considered to be beyond the analyzing power of the present detectors.

The background induced by muons crossing the detector parallel to the iron plates or through gaps between the counters is expected to be negligible. The contribution due to their interactions, which is mainly electromagnetic, is also negligible because the muon intensity itself is so low. The effect due to neutrons produced by the muon nuclear interaction in surrounding rock, is also negligible. If this effect was at all observable, the Phase II detector should have shown 5 times more events compared with Phase I, in proportion to the intensity of muons at the respective depths.

Topological and kinematical limitation on the energy E and momentum P and particle identification are used to select the candidates. The accuracy in the determination of these parameters seems to be at a significant level to cut off the background. Although no special hypothesis for the decay modes has been proposed for the analysis of the observed events, the accepted interpretation for each candidate event seems to be quite unambiguous. The distribution of events in their visible energy is shown in Fig.9. It is not possible to detect any large excess in the number of observed events in the energy region of proton decay. The total number of candidates is about 30 % of the total number of observed events, which are distributed from 200 to 1000 MeV in the visible energy. The number of events per bin is very small with obviously large statistical fluctuations. Furthermore, the numbers estimated from the neutrino flux and their nuclear interaction cross section also have an error of about 30 %. Solar activity during the observation period, Geo-Magnetic latitude effect, neutrino components, interaction characteristics, detector efficiency etc. are some of the many factors to be considered which make it rather difficult to estimate proton decay rate from the rate of observation. One of the nuclear effects to be considered is due to the interaction of mesons produced by nucleon decay inside the parent nucleus. The calculations of Arafune (Ref.13) for the case of iron give the following results;

| no interaction | scattering | charge exchange | absorption |
|---|---|---|---|
| ～40 % | ～15 % | ～13 % | ～30 % |

Pion production is estimated to be about 2 %. Therefore, about 40 % of the nucleon decay signals are completely free from nuclear effects for the case of $^{56}$Fe nucleus. The angular distortion accepted for the candidate events is up to about 30° from the back to back configuration, which is predominantly determined, by Fermi motion within the nucleus. This criterion makes for a clear separation from neutrino induced events, having opening angle of less than 120°. During the measurement on the events, vertices are determined visually on a graph using association of tracks in the two independent views. The particle identification, momentum and energy determination are done in two ways, by ionization and range of the tracks. In any ambiguous case for a track, whether it is a muon or a pion, the interpretation which is closer to the nucleon decay hypothesis has been chosen. Nevertheless, there is no significant bias because the candidate events are very much different from neutrino events in their topological configuration.

The calculation of the nucleon lifetime is based on the total exposure (kty) the number of observed candidates, the expected background and the detection efficiency for a given decay mode, including the nuclear effect. The background is negligible in most of the cases and is significant only for single prong modes. Assuming the nuclear effect of about 40 %, and detection efficiency of 80 %, we estimate the lifetime as follows;

$$\text{lifetime} = 6 \times 10^{32} \times 1.67 \times 0.4 \times 0.8 / 23 = (1.4 \pm 0.5) \times 10^{31} \text{ years} \quad (1)$$

where first two terms give the number of nucleons in 1k ton, 1.67 kty is total exposure factor in this experiment and 23 is number of candidate events, background subtracted. Thus the nucleon lifetime is obtained as $1.4 \times 10^{31}$ years. Partial lifetime for each decay mode can be obtained approximately as it is inversely proportional to the ratio of the number of candidate events for the particular decay mode to the total number. The error is estimated as about $\pm 0.5 \times 10^{31}$ years including statistical and systematic ambiguities.

V. Discussion

There are many theoretical calculations (Ref. 7) for nucleon decay based on the Grand Unification Theory (GUT). Table I shows one of them using the minimal SU(5) model. On the other hand, the Supersymmetry (SUSY) hypothesis has a special emphasis on the decay mode, $\bar{\nu} K^+$, compared with other decay modes. There are many other theories; however, the present experimental status is not sufficiently clear to justify a detailed discussion on the prediction from these theories. Therefore, we discuss here only briefly the two cases mentioned above.

According to the standard model, SU(5) GUT, the proton decay is induced through GUT gauge boson (X-Boson). One of the typical and most abundant modes predicted by this model for the proton decay is to a positron and a neutral pion. The partial lifetime of proton through this channel is about $5 \times 10^{31}$ years from our experiment. According to the theory, the lifetime and Mx are related as follows;

$$t(p \rightarrow e^+ \pi^0) = 1.1 \times 10^{34-36} (M_x/10^{16} \text{ GeV})^4 = 5 \times 10^{31} \text{ years} \quad (2)$$

Substituting the numerical values and taking the central value in the exponent, we get

$$M_x = 1.5 \times 10^{15} \text{ GeV} \quad (3)$$

The value of $M_x$ is not very sensitive to the observed value of the lifetime because it is proportional to the l/4 power of the partial lifetime. This value (3) is one order of magnitude smaller than the value of $2 \times 10^{16}$ GeV estimated from the extrapolation of gauge coupling constants, for strong, week and electromagnetic interactions to higher energy region, where they converge to a common value as predicted by SUSY GUT. (Ref.14) Since there is considerable amount of uncertainty in such a large extrapolation, the value of $M_x$ obtained here may be considered as a reasonable estimate.

The SUSY SU(5) GUT also predicts another possibility for the proton decay through the colored Higgs boson which could decay into SUSY particles and finally go to anti-lepton and anti-quark. The typical decay mode for this process is, $p \rightarrow \bar{\nu} K^+$ with a partial life time of ;

$$t(p \rightarrow \bar{\nu} K^+) = 6.9 \times 10^{27-29} (M_c/10^{16} \text{ GeV})^2 (M_{SUSY}/10^3 \text{ GeV})^2 \text{ years} \quad (4)$$

where $M_c$ and $M_{SUSY}$ are the masses of the colored Higgs boson and SUSY particle respectively. From our result, the partial life time for this channel is about $1.4 \times 10^{31}$ years. A combination of the values say for $M_c$ $1.4 \times 10^{16}$ GeV and $M_{SUSY} = 10^4$ GeV, which gives about the value of $1.4 \times 10^{31}$ years for the lifetime, may be a close solution fitting our observations. Depending on the mass of the SUSY particle, which is expected to be around $10^2 - 10^4$ GeV, the mass of colored Higgs boson can be fixed. Although there are large ambiguities for the SUSY particle, it may be possible to discover it in an accelerator experiment in the future. In any case, as mentioned above, our result seems to be consistent with values expected from a range of minimal standard models of SUSY SU(5) GUT and with the figures discussed by many authors in recent years (Ref.7).

There is another theoretical point of view (Ref.15) that the virtual meson exchange between nucleons may shorten the lifetime of nucleon inside a nucleus. Such an effect may increase the decay probability about 50 % in case of A = 60 nucleons. However, in this case, NN $\rightarrow$ e $\Delta$, $\Delta \rightarrow \pi$ N type of the interaction, the angular distribution of the decay particle will be given by the phase space only. Therefore, it may be difficult to separate these decays from neutrino interactions because of the lack of momentum balance which is an important criterion for detecting ordinary proton decay. It would not be so easy to verify this effect experimentally.

Next, we discuss our results from the point of view of experimental considerations. As mentioned already in the Introduction, there are a number of reports from other experiments (Ref. 11), which have reached a general consensus among themselves that they have not found any conclusive evidence for proton decay event yet and that the life time must be as long as $10^{33}$ years (Ref.12). This conclusion is indeed in direct conflict with our results presented in this paper. In our opinion, there are many points of agreement between the observations in other experiments and ours, as mentioned below.

The Mont Blanc group has reported their results (Ref.16) from their observations with an exposure factor of 0.215 kty, as follows;

$p \rightarrow e^+ \pi^0$ 1 event, $p \rightarrow \bar{\nu} \pi^+$ 6 events and $p \rightarrow \mu^+ K^0$ 2 events

They have observed 9 events out of 21 events which are fully contained. This is consistent with our observations within statistical fluctuations, considering that the $\bar{\nu}\pi^+$ decay mode may have a large background.

The Frejus experiment has observed only one event of type, $p \rightarrow \bar{\nu} K^+$ $K^+ \rightarrow \mu^+ \nu$. According to them, the detection efficiency for this particular type of event is about 8 %. So their observed rate is in reasonable agreement with our result.

One of the results of Kamiokande (Ref.11) shown in Fig.10. The momentum distribution for their events shows two peaks, one at about 240 MeV/c corresponding to muons from kaon decay, and another peak at around 400 MeV/c which correspond to $p \rightarrow \bar{\nu}\pi^+$ if it is analyzed as muon. The second peak may be somewhat broader due to Fermi motion within the oxygen nucleus. The new background rate without these two bins is about half of their estimated value and the number of events after subtraction of the new background in these two bins are 5 and 6 respectively, in their observations with exposure factor of 4.2 kty. Both of these rates are close to our observations in spite of different experimental techniques.

The IMB group has reported at Adelaide Conference (Ref.17), that they have found 4 candidate events for $e^+\pi^0$ during the observation of about 4 kty. Out of these, two events have been rejected because of their association with muon decay signals. The other two events, according to their analysis also have some difficulty to be considered as due to proton decay phenomenon. One of them has a concentrated Cerenkov light cone which may come from a slow proton and the other event has an extra light cone due to a lower energy particle. However, if these features are ascribed to fluctuations in cascade showers, these two events may remain as candidates for proton decay. Assuming such an interpretation, their observed rate of candidate event becomes close to our results.

In general, Cerenkov detectors operating at shallow depths and using anti-coincidence system to remove frequently observed muon events, may be affected in their detection efficiency for cascade showers because of the long tail of the cascade showers. The total number of confined events in the KGF experiments during a running time of 1.67 kty is about 100. A comparison of numbers of these low energy neutrino events between different experiments is not easy because of differences in experimental conditions. Even in the same experiment, low energy events just near the threshold energy are difficult to analyze. In the present paper, such low energy events have not been included in the figure mentioned as about 100 in the previous section.

In this paper, 'proton decay' has been emphasized; as one sees from the results presented above, there is a deficiency in the number of neutron decay candidate events. This may be due to statistical fluctuations or due to a bias against detection of neutron decay from detector efficiency. For instance, $\bar{\nu}\pi^0$ and $\bar{\nu} K^0$ are somewhat difficult to detect because of smaller triggering efficiency and larger background, and $e^+ K^- \rightarrow \bar{\nu}\mu^-$ has no clear topological criterion for easy selection. Only. $e^+\pi^-$ may be good candidate. If one considers only protons, the life time of protons may be adjusted to about half the value from results discussed above as Z/A = 26/56 for the iron nuclei. Concerning neutrons, there is an estimation for nn~ oscillations through which the baryon number is changed by 2. In the KGF experiments, there are 4 candidate events at around 1.5 GeV in visible energy. One of these examples is shown in Fig.11. It is not easy to discuss this phenomenon in greater detail because the annihilation of the anti-neutron within the iron nucleus is very complicated. The period of nn~ oscillation (transition time) estimated on the basis of these candidate events is (5 ± 2) x $10^{31}$ years.

For studies on rare type events deep underground, the main background is due to neutrino interactions which has been discussed by us in our early observation (Ref.18). On this subject, very useful result has been obtained by the Frejus group (Ref.19), using the neutrino beam at CERN. They have reached an important conclusion from this study that, if we consider more than two prong events with energy in the range, 800 - 1000 MeV, and momentum sum less than 400 MeV/c, we should expect a background event rate of 2.5 per kty due to the flux of cosmic ray neutrinos in detectors deep underground. Their results also show that the number of (background) events is very sensitive to the selected ranges of energy and momentum because of their very steep spectrum. If one reduces both the ranges, of energy and momentum, to half of the above values, the expected event rate decreases sharply to about 1/20 of the rate given above. Note that the background of about 0.1 events per kty is small enough to be neglected for the observed rate of proton decay candidates, observed in the KGF experiments and discussed in this paper.

VI. Conclusion

There is experimental evidence for the existence of proton decay phenomenon. Protons decay into a wide spectrum of decay modes. The branching ratios of these modes seem to be of same order of magnitude. However, as a whole, the decay mode with kaon seems to be dominant compared with the decay mode into pion. From a detailed analysis of experimental data for the decay modes; $\bar{\nu}\pi$、$\bar{\nu}K$、$e^+\pi$、$e^+K$, $\mu^+\pi$、$\mu^+K$ : the lifetime of nucleon is estimated as (1.4 ± 0.5) x $10^{31}$ years. This value of lifetime obtained from the 'Phase I and II' experiments at Kolar Gold Fields is well within the range of predictions based upon minimal SU(5) SUSY GUT models.

Finally, it must be mentioned that we assumed, in this paper, the decay modes of simple type for the candidate events to test the existence of nucleon decay. Such an experimental analysis largely depends upon the prediction of theories, because the experiment is not the type for perfect measurement. For example, the sign of the electric charge of the particles is not measured, separation between muons and pions is not clear, and the candidate events which classified into $e^+ K^0$ may partly include the candidate events of $e^+ \eta$. These items are beyond the ability of the present detectors which are the first to be dedicated to find very rare phenomenon like 'Nucleon Decay'.

Appendix

(1) Spatial resolution of the detector

Since the cross-section of the counter is 10 cm x 10 cm, spatial resolution in each observed figure, for independent single counter, is 10 cm.  In case of single counter in a track, ionization of the counter should be considered to trace the track. Hence, the space resolution will be a few cm as shown in Fig.12 b and c.  In case of double counters discharged in a layer for an inclined track, position of the cross point on vertical counter wall is fixed taking into account the ratio of ionization in both the counters. Since the error of ionization is 20% for 10 cm path length, the spatial resolution for the position of the cross point is about 1cm as shown in Fig.12 d.  One can check these spatial resolution in the detector practically in Fig.1 and 2; then one will see that the estimated track is not easy to shift even 1 cm satisfying the ionization of discharged counters.  As a result, most of candidate events have their vertex in an accuracy of about 2 cm error in iron plate where the event originated.

The angular resolution in case of 1 cm x 1 cm counters of 1 cm separation is equal to the case of 10 cm x 10 cm counters with 10 cm separation.  In case of same thickness of iron sheet in between the layers, in above two cases the total size of the observed event is smaller in the former case (high density detector) than the latter (low density detector).  In the analysis of the observed events, required resolution is relative to the size of the observed events.  Therefore, the size of the counter, 10 cm in this experiment, is not very coarse compared with other fine grain detectors.

The ionization, important in the analysis, has an error of 70 $\beta$ $(LP)^{-1/2}$ %, where L is a path length (cm) in the counter and P is a pressure of inner gas in atm.  It comes to about 20 % for 10 cm path length in this experiment.  The ionization is stable within the range of above error in the low energy region of the particle, but it fluctuates more in case of high energy particles due to the energetic knock-on electrons.

Fig.12

(2) The energy of pion in the decay mode P→$\bar{\nu}$ $\pi^+$

The energy of the pion is easily calculated as 479 MeV in case of free proton.  In the case of protons bound in iron nucleus, one must take into account the nuclear effects; binding energy of about 8 MeV, Fermi momentum up to about 250 MeV/c.  In case of proton decay into $\bar{\nu}$ and $\pi^+$, the approximate solution for

$E_{nu} + E_{pi} = M_p - B_a$ and $P_{nu} + P_{pi} = k$ is obtained as follows,

$P_{pi}= [ M_p - B_a - k \cos\phi - (m_{pi}^2 + k^2 \sin^2\phi) /(M_p - B_a - k \cos\phi)] / 2$   (5)

$P_{nu}= [ M_p - B_a + k \cos\phi - (m_{pi}^2 + k^2 \sin^2\phi) /(M_p - B_a - k \cos\phi)] / 2$   (6)

Where $B_a$ is binding energy, k is Fermi momentum and $\phi$ is the angle between k and pion momentum (Fig.13).  The central energy of pion (at $\phi = 90°$) decreases with the increase of k, from 476 MeV for k = 0 to 443 MeV for k = 250 MeV/c.  Assuming the spectrum of k as limited in phase space, isotropic relation for k with respective to the momentum of pion and loss of the events by nuclear interaction, [by exp(-x/L)] the results of the calculation is a distribution at around 460 MeV with a half height width of about 70 MeV.  The observed pions (four cases listed in this paper) are well inside the expected distribution.

Fig.13

(3) Candidate events for P → e$^+$ k$^0$

The three events belonging to this decay mode are shown in Fig. 14 a, b and c. Event No.1766, shown in Fig. 14 a, has a e$^+$ shower of the energy of 370±110 MeV going downwards and has a low energy K$^0$ in the opposite direction which is estimated from its decay into two charged pions with total energy of about 500 MeV. Ev. No. 5038, shown in Fig. 14 b, is also similar to the above, having a low energy K$^0$ decay scheme and e$^+$ shower of energy of 650± 200 MeV in back to back configuration (the energy is probably estimated too high, because one of the counters in the shower shows large ionization). The distance of about 20 cm between shower and K$^0$ decay positions is presumably interpreted as conversion length from K$_L^0$ to K$_S^0$ rather than the fluctuation of the development of e$^+$ shower.

Ev. No. 4910, shown in Fig. 14 c. comprising a few cascade showers, is already listed up in the table of e$^+$ $\pi^0$ mode in this paper. Nevertheless, as it can be seen in the figure, deflection angle, energy balance, wide spread of upward cascade showers are favorable to interpret the event as a candidate for the decay mode P → e$^+$ K$^0$. Dotted lines in the figure are not necessarily gamma rays but showing a wide spread of cascade showers.

Ev. No. 64-125 is shown in Fig.15. It looks a simple event which comprises an electro-magnetic cascade shower going downward and a penetrating particle going upward. Possible explanations for this event are P → e$^+$ K$^0$ or n → e$^+$ K$^-$.

The ambiguity comes from the single counter near the vertex which corresponds to a slow pion or a slow kaon in the above cases. The energy of e-shower is 307±90 MeV, and the energy of the penetrating particle is 310±20 or 265±20 MeV for pion or muon respectively, Although both the cases are possible, a decay electron signal at the end of the track (probably positive charge ), angle between the track and the shower are favorable to interpret the event as the former case, namely P → e$^+$ K$^0$. This choice is also acceptable from the comparison of the event with other candidate events belonging to the same decay mode, for example, Ev. No. 1766 etc. shown in Fig.14. There are many similarities between those events.

Fig.14 a,b,c,d

(4) Probability of the existence of proton decay from the double peaks in the momentum distribution of single track events.

In Fig. 10, the momentum distributions of single tracks treated as muons are shown. In case of KGF, 6 candidate events of kaon decay events without kaon signals and all of 4 events proton decay candidates into antineutrino and pion are included in the distribution. Pion of about 460 MeV is shown as muon of about 400 MeV/c.

In both the experiments, Kamiokande and KGF, there are two peaks in each distribution, one at about 250 MeV/c and another at about 400 MeV/c which correspond to kaon decay and proton decay if the particles in the latter peak are pions. In case of KGF data, 6 events near 250 MeV/c are distributed in two bins as 3 events in each because of errors and 3 events near 400 MeV/c because, out of 4 events, one event is just over 400 MeV/c. Above this momentum region, there is one event at 600 MeV/c in the case of KGF experiment.

The background of the events induced by neutrino interaction assuming a flat spectrum for background, are 3 and 1 per 25 MeV/c bin respectively, taking the average of the number of events in bins except the bins having peaks. The first peak of Kamiokande has 8 events which correspond to about 1 % of probability, if it is due to a fluctuation of the background. Similarly in the second peak near 400 MeV/c about 0.3 % probability is obtained in the independent KGF experiment, as one can see in the figure. The fluctuation of the background is unlikely to be the cause of such double peaks in the momentum distribution, because probability is of the order of $10^{-7}$. These results could be also an evidence for the existence of GUT proton decay, in the decay modes ; P → $\bar{\nu}$ $K^+$ and P → $\bar{\nu}$ $\pi^+$.

(5) The effective exposure factor of the experiments

The exposure factor (kty) is a product of fiducial volume and exposure time, by which the life of proton and neutrino background are estimated. The fiducial volume is defined as follows;

Fid. Vol. = ∫ ∫ $C_i$(x,y,z,l,m,n) $T_i$(x,····n) $A_i$(x,····n) $\rho$ dv dΩ        (7)

Where $C_i$ is probability of confinement of ith type events originating at the position x,y,z with main spatial angle l, m, n, $T_i$ is triggering efficiency of the event and $A_i$ is pick-up efficiency of the analysis of the event. In this paper, it is estimated for the simple and frequent type of the event i.e. decay to anti-neutrino and kaon, which is observed as a kaon decay in the detector. It is clear that the fiducial volume is different for different decay modes, but it is approximated to be the same in the present stage of discussion.

(6) Deflection angle from back to back configuration of proton decay

The deflection angle, $\theta$ in Fig.13, can be calculated from Eq. (5) and (6), using the relation; sin $\theta$ = (k/$P_e$) sin$\phi$ ( neutrino and electron are similar in these calculation ). The maximum angle of $\theta$, assuming the angle $\phi$ is 90° is easily calculated for the cases ; $e^+$ $\pi^0$, $e^+$ $K^0$, $\mu^+$ $\pi^0$, $\mu^+$ $K^0$. For the Fermi momentum k = 200 MeV/c, $\theta_{max}$ is about 25° for the case of anti-lepton and pion, and it is about 35° for the case of anti-lepton and kaon,

This small deflection angle of about 30° makes clear separation of the candidate events from the background neutrino events, because even in case of isotropic distribution, only 6 % of the cases could be in such condition. The neutrino interactions will have somewhat forward-focused secondary particles; then the probability for the above configuration will be less than approximately 1 %. Since the ratio of candidate events to the total confined events observed in the same energy region in this experiment is of the order of 30 %, the actual background to satisfy the particular conditions of proton decay event will be negligible.

Acknowledgements

We are grateful to the Managing Director and staff members of the Bharat Gold Mines Ltd. for their co-operation in providing all facilities for smooth running of the experiment. The Ministry of Science and Education, Japan, is thanked for financial assistance. We are deeply indebted to the scientific and technical staff members of the Tata Institute of Fundamental Research for their help throughout the experiment.